%% file: root_arxiv.tex
\documentclass[letterpaper, 10 pt, conference]{ieeeconf}  

\IEEEoverridecommandlockouts                              

\usepackage{pgfplots}
\pgfplotsset{compat=1.16}
\usepackage{cite}
\usepackage{url}
\usepackage{graphics} 
\usepackage{epsfig} 
\usepackage{mathptmx} 
\usepackage{mathrsfs}
\DeclareMathAlphabet{\mathcal}{OMS}{cmsy}{m}{n}
\usepackage{cuted}

\usepackage{setspace}
\usepackage{amssymb,amsmath,amsfonts}
\usepackage{ntheorem}
\usepackage{booktabs}
\usepackage{makecell}
\let\labelindent\relax
\usepackage{enumitem}
\usepackage{mathtools}
\usepackage{esvect}

\usetikzlibrary{circuits}

\usetikzlibrary{intersections} 

\usetikzlibrary{scopes, arrows, fadings, patterns}

\usetikzlibrary{%
	decorations.pathreplacing,%
	decorations.pathmorphing%
}

\usetikzlibrary{positioning}

\usepackage{bm}
\usepackage{tikz}
\usepackage{pgfplots}
\usepackage{algorithm,algorithmic}  
\usepackage{multirow}
\usepackage{balance}

\newcommand{\Eb}{{\mathbb{E}}}
\newcommand{\Rb}{{\mathbb{R}}}
\newcommand{\Cb}{{\mathbb{C}}}
\newcommand{\Ib}{{\mathbb{I}}}
\newcommand{\Zb}{{\mathbb{Z}}}

\newcommand{\Ac}{{\mathcal{A}}}

\newcommand{\Ic}{{\mathcal{I}}}
\newcommand{\Rc}{{\mathcal{R}}}

\newcommand{\Sc}{{\mathcal{S}}}

\newcommand{\Yc}{{\mathcal{Y}}}

\newcommand{\ra}{{\rightarrow}}
\newcommand{\ift}{{\infty}}
\newcommand{\ls}{\text{ls}}

\DeclareMathOperator{\rs}{{rowspan}}

\DeclareMathOperator{\diag}{{diag}}
\DeclareMathOperator{\supp}{supp}
\DeclareMathOperator{\cov}{Cov}

\theorembodyfont{\normalfont}

\newtheorem{lemma}{\textbf{Lemma}}
\newtheorem{theorem}{\textbf{Theorem}}
\newtheorem{remark}{\textbf{Remark}}
\newtheorem{assumption}{\textbf{Assumption}}
\newtheorem{corollary}{\textbf{Corollary}}

\newtheorem{definition}{\textbf{Definition}}

\title{\LARGE \bf Low Complexity Secure State Estimation Design for Linear System with Non-derogatory Dynamics}

\author{Zishuo Li and Yilin Mo
	\thanks{Zishuo Li and Yilin Mo are with the Department of Automation and BNRist, Tsinghua University, Beijing, China. Email: \{lizs19@mails.tsinghua.edu.cn, ylmo@mail.tsinghua.edu.cn\}. This work is supported by the National Key Research and Development Program of China under Grant 2018AAA0101601.
	}
}

\begin{document}

\maketitle

\begin{abstract}
We consider the problem of estimating the state of a linear Gaussian system in the presence of integrity attacks. The attacker can compromise $p$ out of $m$ sensors, the set of which is fixed and unknown to the system operator, and manipulate the measurements arbitrarily. Under the assumption that all the eigenvalues of system matrix $A$ have geometric multiplicity $1$ ($A$ is non-derogatory), we propose a secure estimation scheme that is resilient to integrity attack as long as the system is $2p$-sparse observable. In the absence of attack, the proposed estimation coincides with Kalman estimation with a certain probability that can be adjusted. Furthermore, our proposed estimator is computational efficient during the security condition checking in the designing phase and during the estimation computing in the online operating phase. A numerical example is provided to corroborate the results and illustrate the performance of the proposed estimator.

\end{abstract}

\section{Introduction}
Cyber-Physical System (CPS) and Internet of Things (IoT) are playing an increasingly important role in critical infrastructures and everyday life. 
The related area of CPS and IoT continues to emerge and expand as costs drop and the confluence of sensors, platforms and networks increases\cite{2018DHS_report}.
Simultaneously, the cyber-security risks and attack surfaces are also increasing \cite{cardenas2008research} since CPS relies on remote sensing devices, communication channels, and spatially distributed processors, which are prone to failures under cyber attacks on the data acquisition and communication channels.
The unintentional faults or malicious attacks could cause severe system damage, economic loss, and environmental degradation, e.g., the Stuxnet launched on Iran’s nuclear facilities\cite{STUXNET}, power blackouts in Brazil \cite{samba_stop} and Ukraine \cite{Ukraine_Blackout}, etc. The research community has recognized the importance of CPS security, especially the design of secure detection, estimation, and control strategy\cite{cardenas2009challenges}. 

Recently, substantial research efforts have been devoted to secure state estimation against malicious sensors.
One of the research paths is the finite horizon approach, where only measurements in a finite time-window are considered when recovering the system state. 
The problem is combinatorial in nature since searching for the corrupted sensors involves minimizing the $\ell_0$ norm \cite{FawziTAC2014}, which is an NP problem.
Fawzi et al.~\cite{FawziTAC2014} propose an algorithm using its convex relaxation, i.e., $\ell_1$ optimization, to solve the problem in the absence of noise. 
Similarly, Shoukry and Tabuada \cite{ShoukryTAC2016} adopt a $2$-norm batch optimization approach to solve the state estimation problem.
Shoukry et al.~\cite{Shoukry2017} propose an algorithm in the presence of noise by searching for reliable sensors using consistency check, and the searching complexity is reduced based on Satisfiability Modulo Theory.
However, in these works, the sensory data out of the time window are discarded, which may cause performance degradation and estimation delay.

Another solution is the switch estimator\cite{Mishra2017TCNS}\cite{yorie}\cite{luTAC2019} where multiple estimates are maintained based on measurements from different subsets of sensors, and the system operator switches between these estimates based on the evaluation of their reliability by consistency checking or malicious detection algorithms. However, the combinatorial nature of estimation candidates or sensor subsets may incur heavy computation and storage burden of the devices.
In view of these problems, Liu et al. \cite{liuxinghua-TAC2020} design local estimators whose weighted sum coincides with the Kalman estimate with a certain probability in the absence of attack. The local estimates are fused securely by a quadratic programming problem with an $\ell_1$ term to handle the sparse outliers.

Even though there are efficient algorithms computing the estimation by introducing $\ell_1$ relaxation, the design of the secure state estimation is still computational challenging. For secure state recovery problem in the presence of $p$ malicious sensors, it is required that the system needs to be $2p$-sparse observable\cite{ShoukryTAC2016}, and calculating the sparse observability index for general systems is proved to be NP-hard\cite{mao2021computational}. Besides sparse observability, other results \cite{FawziTAC2014}\cite{liuxinghua-TAC2020}\cite{sandberg_TAC2014} impose stronger conditions for system resiliency, whose validations are also NP problems.
However, it has been observed by Mao et al. \cite{mao2021computational} that when the eigenvalues of the system matrix $A$ have unitary multiplicity ($A$ is non-derogatory), the computational complexity of secure state reconstruction without noise can be significantly reduced.
Leveraging upon this assumption, we propose our secure estimation scheme in the formulation of LTI system with Gaussian noise, and it has the following merits:
\begin{itemize}[left=0pt]
	\item In the presence of $p$ compromised sensors, the proposed estimation is secure if the system is $2p$-sparse observable. In the absence of attack, the proposed estimation coincides with Kalman estimation with certain probability, which can be adjusted.
	\item During the designing phase, the sparse observability index can be computed with low complexity. During the algorithm operating phase, the proposed estimation is formulated as the solution of a convex optimization problem which can be computed efficiently.
\end{itemize}

\textit{Organization:} We introduce the problem formulation and preliminary results in Section \ref{sec:problem}. The main results are provided in Section \ref{sec:main_result} and collaborated by numerical simulation in Section \ref{sec:sim}. Section \ref{sec:conclusion} finally concludes the paper.

\textit{Notations:}
Cardinality of a set $\Sc$ is denoted as $|\Sc|$. Conjugate transpose of a matrix $A$ is denoted as $A'$.
The determinant of a matrix is represented by $\det(\cdot)$. 
Diagonal matrix with diagonal elements $A_1,\cdots,A_k$ is denoted as $\text{diag}(A_1,\cdots,A_k)$.
Denote the span of row vectors of matrix $A$ as $\rs(A)$.
All-one vector with size $m\times 1$ is denoted as $\mathbf{1}_{m}$. 
$\Rb^n$ is the space of real vectors with $n$ entries. 
The set of complex matrices with $m$ rows and $n$ columns is denoted as $\Cb^{m\times n}$.

\section{Problem Formulation and Preliminary Results}\label{sec:problem}	
\subsection{Secure dynamic state estimation}
In this paper, we consider the linear time-invariant system with Gaussian noise:
\begin{align}
x(k+1)&=A x(k)+w(k) , \label{eq:system} \\
y(k)&=C x(k)+v(k)+a(k) ,\label{eq:y_i_def}
\end{align}
where $x(k) \in \mathbb{R}^{n}$ is the system state, $w(k) \sim {N}(0, Q)$ and $v(k) \sim {N}(0, R)$ are i.i.d. Gaussian process noise and measurement noise with zero mean and covariance matrix $Q$ and $R$.  
The vector $y(k)\in \mathbb{R}^{m}$ is the collection of measurements from all $m$ sensors, and $i$-th entry $y_i(k)$ is the measurement from sensor $i$.
The vector $a(k)$ denotes the bias injected by an adversary and $a_i(k)$ is the attack on sensor $i$. Define $$z(k)=C x(k)+v(k)$$ as the true measurements without the attack.
The initial state $x(0) \sim {N}(0, \Sigma)$ is assumed to be zero mean Gaussian random variable and is independent from the process noise $\{w(k)\}$.

The secure dynamic estimation problem aims at recovering system state $x(k)$ at every time $k$ based on all history observations $\{y(t),0\leq t\leq k\}$ which have been partly manipulated by the malicious attacker.
It is conventional in the literature \cite{FawziTAC2014}\cite{Shoukry2017} that the attacker can only compromise a fixed subset of sensors with known maximum cardinality. 
Denote the index set of all sensors as $\Rc \triangleq\{1,2, \ldots, m\}$. 
Define the support of vector $a\in\Rb^{n}$ as $\supp(a)\triangleq \left\{i| 1\leq i\leq n , a_i\neq0 \right\}$ where $a_i$ is the $i$-th entry of vector $a$.
We have the following assumptions on the malicious adversary. 

\begin{definition}[Sparse Attack]\label{def:attack}
	The attack is called a $p$-sparse attack if the vector sequence $a(k)$ satisfies that,
	there exists a time invariant index set $\Ic\subseteq \Rc $ with $|\Ic| = p$ such that $\bigcup_{k=1}^{\infty} \supp\left\{a(k)\right\} = \Ic$.
\end{definition}

Closely related to the sparse attack, we introduce the notion of sparse observability that characterizes the system observability in the presence of attack.

\begin{definition}[Sparse observability]\label{df:sparse_obs}
	The sparse observability index of system \eqref{eq:system}-\eqref{eq:y_i_def} is the largest integer $s$ such that system\footnote{The matrix $C_{\Rc\setminus\Ic}$ represents the matrix composed of rows of $C$ with row index in $\Rc\setminus\Ic$.} $(A,C_{\Rc\setminus\Ic})$ is observable for any set of sensors $\Ic\subset\Rc$ with cardinality $|\Ic| = s$. When the sparse observability index is $s$, we say that the system with pair $(A,C)$ is $s$-sparse observable.
\end{definition}
Define $y(k_1:k_2)$ as the sequence $\{y(k_1),y(k_1+1),\cdots,y(k_2)\}$. Similar notation is also applied on $z(k)$.	
For linear Gaussian noise system, the estimation is secure if the estimation error is bounded by a constant term irrelevant to the attack. 

\begin{definition}[Secure estimator]\label{def:resi}
	An estimator is an infinite sequence of mappings $g=\{g_k\}_{k=1}^{\ift}$ where $g_k$ is a mapping from all the history observations to an estimation at time $k$:
	$$g_k\left(y(0:k)\right)=\hat{x}(k).$$
	Define the estimation difference introduced by attack as
	$$d_k\triangleq \left\|g_k\left(z(0:k)\right)-g_k\left(y(0:k)\right) \right\|_2 .
	$$
	The estimator is said to be secure against the $p$-sparse attack if the following holds:
	$$\sup_{k\in\Zb^+} \Eb \left[d_k^2\right] < \ift ,$$
	where $\Eb$ is the expectation with respect to the probability measure generated by the Gaussian noise $\{w(k)\}$ and $\{v(k)\}$.
\end{definition}
If all sensors are benign, i.e., $a(k)=\mathbf{0}$ for all $k$, the optimal state estimator is the classical Kalman filter:
\begin{align*}
\hat{x}(k)&=\hat{x}(k | k-1)+K(k)\left[y(k)-C \hat{x}(k  | k-1)\right] ,\\
P(k)&=P(k  | k-1)-K(k) C P(k  | k-1),
\end{align*}
where
\begin{align*}
&\hat{x}(k  | k-1)=A \hat{x}(k-1), P(k  | k-1)=A P(k-1) A{'}+Q ,\\	
&K(k)=P(k  | k-1) C{'}\left(C P(k  | k-1) C{'}+R\right)^{-1},
\end{align*}
with initial condition $\hat{x}(0  |-1)=0,\ P(0  |-1)=\Sigma $.
It is well-known that for observable system, the estimation error covariance matrices $P(k)$ and the gain $K(k)$ will converge to the following matrices $P$ and $K$:
\begin{align*}
&P \triangleq \lim _{k \rightarrow \infty} P(k),\ P_{+}=A P A{'}+Q ,\\
&K \triangleq P_{+} C{'}\left(C P_{+} C{'}+R\right)^{-1}.
\end{align*}
Since typically the control system will be running for an extended period of time, we focus on the case where the Kalman filter is in steady state, and thus the Kalman filter reduces to the following fixed-gain linear estimator:
\begin{equation}\label{eq:fix_gain_kalman}
\hat{x}(k+1)=(A-K C A) \hat{x}(k)+K y(k+1) .
\end{equation}
Before introducing our work, we first recall some results in the previous work that decomposes the fix gain Kalman filter to local estimates and recovers it by an optimization problem.

\subsection{Preliminary Results}\label{sec:preli}
The following preliminary results in this subsection are from \cite{liuxinghua-TAC2020}.
We introduce the following assumption:
\begin{assumption}\label{as:distinct_eigvalue}
	$A-K C A$ has $n$ distinct eigenvalues. Moreover, $A-K C A$ and $A$ do not share any eigenvalue.
\end{assumption}
Since $A-K C A$ has distinct eigenvalues, it can be diagonalized as:
\begin{equation}\label{eq:VLambda}
A-K C A=V \Pi V^{-1},
\end{equation}
where $\Pi$ is a diagonal matrix with the eigenvalues of $A-KCA$ as diagonal entries. Denote these $n$ eigenvalues (diagonal entries) as $\pi_{1},\cdots,\pi_{n}$. 
We design local estimation $\zeta_{i}(k)$ with the following dynamic:
\begin{equation}\label{eq:def_zeta}
\zeta_{i}(k+1)=\Pi \zeta_{i}(k)+\mathbf{1}_{n} y_{i}(k+1) .
\end{equation}
In other words, $\zeta_{i}(k)$ is the state of a estimator with dynamics similar as $A-KCA$ while only takes observations from sensor $i$.

Define $G_i$ as
\begin{equation}\label{eq:def_Gi}
G_{i} \triangleq\left[\begin{array}{c}
C_{i} A\left(A-\pi_{1} I\right)^{-1} \\
\vdots \\
C_{i} A\left(A-\pi_{n} I\right)^{-1}
\end{array}\right].
\end{equation}
In the following, we claim that local estimate $\zeta_i(k)$ is actually a stable estimate of state $x(k)$ transformed by the constant matrix $G_i$, i.e., their difference $\epsilon_i(k)\triangleq\zeta_{i}(k)-G_ix(k)$ is a stationary Gaussian process.
The following result has been proved in \cite{liuxinghua-TAC2020}. 
\begin{lemma}[\hspace{-0.0001pt}\cite{liuxinghua-TAC2020}]\label{lm:epsilon}
	Let $\epsilon_i(k)\triangleq\zeta_{i}(k)-G_ix(k)$, then 
	\begin{align}
	\epsilon_{i}(k+1)= \Pi \epsilon_{i}(k)+\left(G_{i}-\mathbf{1}_{n} C_{i}\right) w(k) \notag \\
	-\mathbf{1}_{n} v_{i}(k+1)&-\mathbf{1}_{n} a_{i}(k+1) . \label{eq:epsilon}
	\end{align}
\end{lemma}
%
Noticing that $\Pi$ is a strictly stable matrix (all eigenvalues are within the open unit disk) and $w(k),v(k)$ are Guaissain zero-mean noise, when the attack is absence, i.e., $a_i(k)=0$, the residue $\epsilon_i(k)$ is a stationary Gaussian process.
Define $\tilde{Q} \in \Cb^{m n \times m n}$ as the covariance of noise term $\left(G_{i}-\mathbf{1}_{n} C_{i}\right) w(k) -\mathbf{1}_{n} v_{i}(k+1)$ for all $i$, i.e.,
\begin{align*}
\tilde{Q} \triangleq&
\cov\left(	\begin{bmatrix}
G_1-\mathbf{1}_{n} C_1 \\
\vdots \\
G_m-\mathbf{1}_{n} C_m
\end{bmatrix} w(k)\right)+
\cov\left(	\begin{bmatrix}
\mathbf{1}_{n} v_{1}(k+1)\\
\vdots \\
\mathbf{1}_{n} v_{m}(k+1)
\end{bmatrix} \right)
\\
=&
\begin{bmatrix}
G_1-\mathbf{1}_{n} C_1 \\
\vdots \\
G_m-\mathbf{1}_{n} C_m
\end{bmatrix}
Q\begin{bmatrix}
G_1-\mathbf{1}_{n} C_1 \\
\vdots \\
G_m-\mathbf{1}_{n} C_m
\end{bmatrix}^{'}
+ R\otimes \mathbf{1}_{n\times n},
\end{align*}
where $\otimes$ is the Kronecker product.
Define $\tilde{\Pi}\in \Cb^{m n \times m n}$ as
$$
\tilde{\Pi} \triangleq\left[\begin{array}{ccc}
\Pi & & \\
& \ddots & \\
& & \Pi
\end{array}\right].
$$
As $k\ra\ift$, the stable covariance of $\epsilon(k)\triangleq\left[\epsilon_1(k){'},\cdots,\epsilon_m(k){'}\right]{'}$ is the solution $\tilde{W}$ of the following Lyapunov equation:
$$
\tilde{W}=\tilde{\Pi} \tilde{W} \tilde{\Pi}{'}+\tilde{Q}.
$$ 
The matrix $\tilde{W}$ is well-defined since $\Pi$ is strictly stable. As a result, the secure estimation can be recovered by the solution of the following optimization problem where $\zeta(k)\triangleq\left[\zeta_1(k){'},\cdots,\zeta_m(k){'}\right]{'} $ and $G\triangleq\left[G{'}_1,\cdots,G{'}_m\right]{'} $.
\begin{subequations}\label{pb:old_lasso}
	\begin{align}
	&\underset{\check{x}(k), \mu(k), \nu(k)}{\operatorname{minimize}}\quad \frac{1}{2} \mu(k){'} \tilde{W}^{-1} \mu(k) + \gamma \|\nu(k)\|_1 \label{min:old_lasso} \\
	& \text{ subject to}\quad
	\zeta(k)=
	G \check{x}(k)+\mu(k)+\nu(k). \label{eq:old_lasso}
	\end{align}
\end{subequations}
The parameter $\gamma$ is a non-negative constant chosen by the system operator.
According to \cite{liuxinghua-TAC2020}, the solution $\check{x}(k)$ to problem (\ref{pb:old_lasso}) is a secure estimation and has the following properties.
\begin{theorem}[\hspace{-0.01pt}\cite{liuxinghua-TAC2020} Theorem 4]\label{th:TAC}
	In the presence of $p$-sparse attack, the state estimation $\check{x}(k)$ solved from problem \eqref{pb:old_lasso} is secure if the following inequality holds for all $u \neq \mathbf{0}$, $u\in\Rb^n$:
	\begin{equation}\label{eq:cond}
	\sum_{i \in \mathcal{I}}\left\|G_{i} u\right\|_{1}<\sum_{i \in \mathcal{I}^{c}}\left\|G_{i} u\right\|_{1}, \quad \forall\ \Ic\subset \Rc, |\mathcal{I}|\leq p .
	\end{equation}
\end{theorem}
Even though Theorem \ref{th:TAC} establishes the sufficient condition of the estimation to be secure, validating \eqref{eq:cond} is NP-hard.
In the following section, we reduce the complexity of checking condition \eqref{eq:cond} by transforming $G_i$ to its canonical form under the assumption that the geometric multiplicities of all the eigenvalues of $A$ are $1$ ($A$ is non-derogatory).

\section{Main Results}\label{sec:main_result}
In this section, under the assumption on $A$, we first prove that the span of rows of $G_i$ coincides with the observability space of $(A,C_i)$, which implies that the matrix $G_i$ has a canonical form under row operations. Based on the canonical form, a new secure estimation scheme is proposed.  

\subsection{Canonical form of $G_i$}\label{subsec:transform}
In order to prevent degeneration problem, we introduce the following assumption.
\begin{assumption}\label{as:geo_mul_1}
	System matrix $A$ is non-singular, and all the eigenvalues of $A$ have geometric multiplicity $1$.
\end{assumption}
Without loss of generality, based on Assumption \ref{as:geo_mul_1}, we assume that $A$ is in the following Jordan canonical form:
\begin{align*}
A=\begin{bmatrix}
J_{1} & \mathbf{0} & \cdots & \mathbf{0} \\
\mathbf{0} & J_{2} & \cdots & \mathbf{0} \\
\vdots & \vdots & \ddots & \vdots \\
\mathbf{0} & \mathbf{0} & \cdots & J_{l} 
\end{bmatrix},\
J_k=
\begin{bmatrix}
\lambda_{k} & 1 & 0 &  \cdots & {0} \\
{0} & \lambda_{k} & 1 & \cdots & {0} \\
\vdots & \ddots & \ddots   & \ddots & \vdots \\
\vdots & \ddots & \ddots  & \ddots & 1 \\
{0} & \cdots & \cdots & 0 & \lambda_{k} 
\end{bmatrix},
\end{align*}
and the (complex) eigenvalues with respect to distinct blocks are different, i.e., $k_1\neq k_2$ implies $\lambda_{k_1}\neq\lambda_{k_2}$. 
Define the observability matrix of system $(A,C_i)$ as 
\begin{equation}\label{eq:def_O}
O_{i} \triangleq\left[\begin{array}{c|c|c|c}
C_{i}{'} &
\left(C_{i} A\right){'} &
\cdots &
\left(C_{i} A^{n-1}\right){'}
\end{array}\right]{'}.
\end{equation}
Before continuing on, we need the following notation of state observability. 
Define $\Sc_j$ as the index set of sensors that can observe state $j$, i.e.
\begin{equation}\label{eq:def_Oc}
\Sc_j\triangleq \{i\in\Rc\ |\ O_i{'} e_j\neq \mathbf{0} \},
\end{equation}
where $\Rc\triangleq \{1,2,\cdots,m\}$ is the index set of all sensors and $e_j$ is the $n$-dimensional canonical basis vector with 1 on the $j$-th entry and 0 on the other entries.
We have the following theorem characterizing the structure of $G_i$. 
\begin{theorem}\label{th:span}
	Assume system matrix $A$ satisfies Assumption \ref{as:geo_mul_1}, then the following equation holds:
	\begin{align}
	\rs(G_i)=\rs(O_i)=\rs(H_i) ,
	\end{align}
	where $H_i$ is the following diagonal matrix
	\begin{equation*}
	H_i\triangleq \begin{bmatrix}
	\Ib_{i\in\Sc_1} & & & \\
	&\Ib_{i\in\Sc_2} & &  \\
	& & \ddots &  \\
	& & & \Ib_{i\in\Sc_n}
	\end{bmatrix} ,
	\end{equation*}
	and $\mathbb{I}_{i\in\Sc}$ is the indicator function that takes the value 1 when ${i\in\Sc}$ and value 0 when ${i\notin\Sc}$.
\end{theorem}

\begin{proof}
	Define the characteristic polynomial of $A$ as $p(x)=a_n x^n +\cdots+a_1 x +a_0$.
	Define polynomial fraction  $q_\pi(x)$ with respect to constant $\pi$ as
	$q_\pi(x)\triangleq\frac{p(x)-p(\pi)}{x-\pi}$ where $x\neq \pi$.
	For polynomial fraction $q_\pi(x)$, we have
	\begin{align*}
	&\frac{p(x)-p(\pi)}{x-\pi}\\
	=&\frac{1}{x-\pi}\left(a_n(x^n-\pi^n)+a_{n-1}(x^{n-1}-\pi^{n-1})+\cdots+a_1(x-\pi)\right)\\
	=&a_n\left(x^{n-1}+\pi x^{n-2}+\pi^2x^{n-3}+\cdots+\pi^{n-1}\right)\\
	&+a_{n-1}\left(x^{n-2}+\pi x^{n-3}+\pi^2x^{n-4}+\cdots+\pi^{n-2}\right)\\
	&+\cdots \\
	&+a_{2}\left(x+\pi\right)\\
	&+a_1,
	\end{align*}
	where the last inequality comes from the fact that 
	$$\frac{x^n-\pi^n}{x-\pi}=\pi^0x^{n-1}+\pi^1 x^{n-2}+\pi^2x^{n-3}+\cdots+\pi^{n-1}x^0.$$
	The above interpretation of $q_\pi(x)$ is also valid when $x$ is a square matrix.
	As a result, by rearranging the terms of $A$ with the same power, one obtains
	\begin{align*}
	q_{\pi_j}(A)=&a_n\left(A^{n-1}+\pi_j A^{n-2}+\cdots+\pi_j^{n-1}I\right)\\
	&+a_{n-1}\left(A^{n-2}+\pi_j A^{n-3}+\cdots+\pi_j^{n-2}I\right)\\
	&+\cdots+a_{2}\left(A+\pi_j I\right)+\cdots+a_1 I\\
	=&a_nA^{n-1}+\left(a_n\pi_j+a_{n-1}\right)A^{n-2}+\cdots\\
	&+\left(a_n\pi_j^{n-1}+a_{n-1}\pi_j^{n-2}+\cdots+a_2\pi_j+a_1\right)I.
	\end{align*}
	Define $b_{j,k}$ as the constant scalar coefficient of $A^k$ in $q_{\pi_j}(A)$:
	\begin{equation}\label{eq:bjk}
	b_{j,k}\triangleq\sum_{i=0}^{n-k-1} a_{i+k+1} \pi_j^i.
	\end{equation}
	Thus, $q_{\pi_j}(A)=\sum_{k=0}^{n-1}b_{j,k}A^{k}$.
	The $j$-th row of matrix $G_i$ can be reformulated as
	\begin{align*}
	&C_{i} A\left(A-\pi_{j} I\right)^{-1}\notag \\
	=&-\frac{1}{p(\pi_j)} C_i A q_{\pi_j}(A)\notag \\
	=&-\frac{1}{p(\pi_j)} C_i \left(\sum_{k=0}^{n-1}b_{j,k}A^{k+1}\right)\notag \\
	=&-\frac{1}{p(\pi_j)}\begin{bmatrix}
	b_{j,0} & b_{j,1} & \cdots  & b_{j,n-1} 
	\end{bmatrix} 
	\begin{bmatrix}
	C_i A \\
	C_i A^2 \\
	\vdots \\
	C_i A^n
	\end{bmatrix}\notag \\
	=&-\frac{1}{p(\pi_j)}\begin{bmatrix}
	b_{j,0} & b_{j,1} & \cdots  & b_{j,n-1} 
	\end{bmatrix} O_i A. 
	\end{align*}

	Therefore, $G_i$ can be interpreted as follows
	\begin{align}\label{eq:Gi=DOA}
	G_i = \mathcal{D}_1\begin{bmatrix}
	b_{1,0} & b_{1,1} & \cdots  & b_{1,n-1} \\
	b_{2,0} & b_{2,1} & \cdots  & b_{2,n-1} \\
	\vdots & \vdots & \ddots  & \vdots \\
	b_{n,0} & b_{n,1} & \cdots  & b_{n,n-1} 
	\end{bmatrix}
	O_i A 
	= & \mathcal{D}_1\mathcal{D}_2\mathcal{D}_3 O_i A , 
	\end{align}
	where $\mathcal{D}_1\triangleq\text{diag}\left(-\frac{1}{p(\pi_1)},-\frac{1}{p(\pi_2)},\cdots,-\frac{1}{p(\pi_n)}\right)$,
	\begin{align*}
	\mathcal{D}_2\triangleq\hspace{-2pt}
	\begin{bmatrix}
	\pi_1^{n-1} & \pi_1^{n-2} & \cdots  & 1 \\
	\pi_2^{n-1} & \pi_2^{n-2} & \cdots  & 1 \\
	\vdots & \vdots & \cdots  & \vdots \\
	\pi_n^{n-1} & \pi_n^{n-2} & \cdots  & 1
	\end{bmatrix}, 
	\mathcal{D}_3\triangleq\hspace{-2pt}
	\begin{bmatrix}
	a_n & 0 & \cdots &   0 \\
	a_{n-1} & a_n & \cdots &   0 \\
	\vdots & \vdots & \ddots  & \vdots \\
	a_1 & a_2 & \cdots  & a_n 
	\end{bmatrix}.
	\end{align*}
	According to Assumption \ref{as:distinct_eigvalue}, all $\pi_j$ are distinct eigenvalues and they are not the eigenvalues of $A$, i.e. the diagonal matrix $\mathcal{D}_1$ and the Vandermonde matrix $\mathcal{D}_2$ are invertible. Moreover, $a_n=1$. Therefore, the lower triangular Toeplitz matrix $\mathcal{D}_3$ is invertible and thus $\rs(G_i)=\rs(O_i A)$ from equation \eqref{eq:Gi=DOA}. 		
	We continue to prove $\rs(O_i)=\rs(O_i A)$. Considering that $A^n=-a_{n-1}A^{n-1}-\cdots-a_0 I$, one obtains the following equation \eqref{eq:O_OA}.
	\begin{equation}
	\label{eq:O_OA}
	O_i A=
	\begin{bmatrix}
	0 & 1 & 0 &  \cdots & 0 \\
	0 & 0 & 1 &  \cdots & 0 \\
	\vdots & \vdots & \vdots & \ddots & \vdots \\
	0 & 0 & 0 &  \cdots & 1 \\
	-a_0 & -a_1 & -a_2 & \cdots &  -a_{n-1}
	\end{bmatrix}
	O_i .
	\end{equation}	
	According to Assumption \ref{as:geo_mul_1}, $A$ is invertible and $a_0=(-1)^n\det(A)\neq 0$, which leads to the equation that $\rs(O_i)=\rs(O_i A)$.
	Since $A$ is assumed to be in the Jordan canonical form and all eigenvalues have geometric multiplicity 1, one can verify that nonzero columns of $O_i$ are linear independent, and equivalently nonzero columns of $G_i$ are linear independent. Therefore, $i\in\Sc_j$ is equivalent to that $j$-th column of $G_i$ is non-zero, i.e., $G_i$ has the same row-span with the canonical form $H_i$.
\end{proof}

Recall that $\rs(\cdot)$ represents the span of row vectors of a matrix.
According to Theorem \ref{th:span} and definition of $\rs(\cdot)$, one directly obtains the following corollary:
\begin{corollary}\label{co:P_i}
	For every sensor index $i\in\Rc$, there exists an invertible matrix $P_i$ such that $P_iG_i=H_i$.
\end{corollary}

After transformation $P_i$, matrix $G_i$ is transformed into canonical form $H_{i}$ whose rows are either canonical basis vectors or zero vectors. 
The non-zero entries of $H_i$ records the state observability of sensor $i$. Therefore, the sparse observability index can be directly obtained from diagonal matrices $H_i,i\in\{1,2,\cdots,m\}$ or, equivalently, from sets $\Sc_j,j\in\{1,2,\cdots,n\}$.
\begin{corollary}\label{co:sparse_obs}
	Denote $s$ as the sparse observability index of system \eqref{eq:system}-\eqref{eq:y_i_def}. Then 
	\begin{equation*}
	s=\min_{j\in\{1,2,\cdots,n\}} |\Sc_{j}| - 1 .
	\end{equation*}
\end{corollary} 
\begin{proof}
	According to the definition of $s$, for arbitrary $\overline{s}$ that satisfies $\overline{s}\geq s+1$, there exists a state index $j^*$ and a sensor index set $\Ic^*$ with $|\Ic^*|=\overline{s}$ such that $\Sc_{j^*} \cap \left(\Rc\setminus\Ic^*\right)=\varnothing$.
	As a result, state $j^*$ can not be observed by any sensor in $\Rc\setminus \Ic^*$, i.e.,
	\begin{equation*}
	e_{j^*}\notin \rs(O_i),\ \forall i\in\Rc\setminus \Ic^*.
	\end{equation*}
	and thus system $(A,C_{\Rc\setminus\Ic^*})$ is not observable.
	For arbitrary $\underline{s}$ that satisfies $\underline{s}\leq s$, arbitrary $j$ and arbitrary $\Ic$ with $|\Ic|=\underline{s}$, one obtains $\Sc_{j} \cap \left(\Rc\setminus\Ic\right)\neq\varnothing$, which means for all $j$, there exists $i^*\in\Rc\setminus \Ic$ such that: $e_{j}\in \rs(O_{i^*})$. Therefore, system $(A,C_{\Rc\setminus\Ic})$ is observable. According to Definition \ref{df:sparse_obs}, the system is $s$-sparse observable.
\end{proof}

In view of Corollary \ref{co:sparse_obs}, the system sparse observability index can be obtained with low computational complexity since calculating cardinality of $\Sc_j$ can be done in \emph{polynomial time} with respect to $m$ and $n$.
In the following subsection, we will propose an information fusion scheme that is secure in the presence of $p$-sparse attack as long as the system is $2p$-sparse observable.

\subsection{Secure Information Fusion}
Recalling the transformation $P_i$ introduced in Corollary \ref{co:P_i}, define $\tilde{P} \triangleq \text{diag}\left(P_1,\cdots,P_m\right),\ \tilde{M}\triangleq\tilde{P}\tilde{W}\tilde{P}{'}$ and
\begin{align}
\Yc (k)\triangleq
\begin{bmatrix}
P_1\zeta_{1}(k) \\
\vdots \\
P_m\zeta_{m}(k)
\end{bmatrix}, \
H\triangleq\begin{bmatrix}
H_{1} \\
\vdots \\
H_{m}
\end{bmatrix} .	
\end{align}
We present the following optimization problem whose solution is our proposed estimation. The constant scalar $\gamma$ is an adjustable parameter.
\begin{subequations}\label{pb:lasso}
	\begin{align}
	&\underset{\tilde{x}(k), \mu(k),\nu(k)}{\operatorname{minimize}}\quad \frac{1}{2} \mu(k){'} \tilde{M}^{-1} \mu(k) + \gamma\|\nu(k)\|_1  \\
	&\ \text{subject to }\quad
	\Yc(k)=H \tilde{x}(k) +\mu(k)+\nu(k). 
	\end{align}
\end{subequations}

We first prove that, in the absence of attack, the estimation  $\tilde{x}(k)$ is optimal for certain probability that can be adjusted.
We need to define the following matrices to facilitate the understanding of the result. 
Define
$$F_i\triangleq V\diag(V^{-1}K_i),\ F=\begin{bmatrix} F_1&F_2&\cdots & F_m \end{bmatrix},$$
where $V$ is defined in \eqref{eq:VLambda} and $K_i$ is the $i$-th column of Kalman gain $K$. Moreover, $\diag(V^{-1}K_i)$ is an $n\times n$ diagonal matrix with the $j$-th diagonal entry equals to $j$-th element of vector $V^{-1}K_i$.
Recalling that $\epsilon(k)\triangleq\left[\epsilon_1(k)',\cdots,\epsilon_m(k)'\right]'$ and $\epsilon_i(k)=\zeta_i(k)-G_ix(k)$ from Lemma \ref{lm:epsilon},
the following theorem characterizes the estimation $\tilde{x}(k)$ in the absence of attack.

\begin{theorem}\label{th:no_attack}
	In the absence of attack, if the parameter $\gamma$ in problem (\ref{pb:lasso}) satisfies
	\begin{align}\label{eq:kalman_cond}
	\left\|\tilde{M}^{-1}\left(I-GF\right)\epsilon(k)\right\|_\infty \leq \gamma,
	\end{align}
	then our proposed estimation $\tilde{x}(k)$ solved from problem \eqref{pb:lasso} is equivalent to the Kalman estimation $\hat{x}(k)$ defined in (\ref{eq:fix_gain_kalman}), i.e.,
	\begin{equation}\label{eq:eq_to_kalman}
	\tilde{x}(k)=\hat{x}(k).
	\end{equation}
\end{theorem}
\begin{remark}
	Noticing that $\epsilon(k)$ is a stationary Gaussian process from Lemma \ref{lm:epsilon}, the probability that \eqref{eq:kalman_cond} holds can be calculated given the parameters $A,C,Q,R$ of the linear Gaussian system \eqref{eq:system}-\eqref{eq:y_i_def}.
	By tuning design parameter $\gamma$, the probability of recovering the Kalman estimation can be adjusted.
\end{remark}
\begin{proof}
	Consider the following least square problem by suppressing $\nu(k)=\mathbf{0}$ in problem \eqref{pb:lasso}.
	\begin{subequations}\label{pb:least_square}
		\begin{align}
		&\underset{\tilde{x}_\ls (k), \varphi(k)}{\operatorname{minimize}}\quad\ \frac{1}{2} \varphi(k)' \tilde{M}^{-1} \varphi(k)   \\
		&\text{subject to }\quad
		\Yc(k)=H \tilde{x}_\ls(k) +\varphi(k). 
		\end{align}
	\end{subequations}
	It can be verified that the solution is equivalent to problem (30) in \cite{liuxinghua-TAC2020}.
	Thus, the solution $\varphi(k)$ of problem \eqref{pb:least_square} satisfies the following equation due to Theorem 1 in \cite{liuxinghua-TAC2020}:
	\begin{equation}\label{eq:ls_recover}
	\tilde{x}_\ls(k)=\hat{x}(k),\ \varphi(k)=(I-GF)\epsilon(k),
	\end{equation}
	where $\hat{x}(k)$ is the fixed gain Kalman estimation defined in \eqref{eq:fix_gain_kalman}.
	Considering the KKT condition of problem \eqref{pb:lasso}, one obtains that if $\left\|\tilde{M}^{-1} \varphi(k)\right\|_\ift\leq\gamma$, then the solution $\tilde{x}(k),\mu(k),\nu(k)$ satisfy
	\begin{equation}\label{eq:lasso_recover}
	\tilde{x}(k)=\tilde{x}_\ls(k)=\hat{x}(k),\ \mu(k)=\varphi(k),\ \nu(k)=\mathbf{0}.
	\end{equation}
	Combining \eqref{eq:ls_recover} and \eqref{eq:lasso_recover}, result in Theorem \ref{th:no_attack} is obtained.
\end{proof}

In the presence of $p$-sparse attack, 
we have the following theorem demonstrating that the estimation $\tilde{x}(k)$ solved from problem \eqref{pb:lasso} is secure. 

\begin{theorem}\label{th:main}
	In the presence of arbitrary $p$-sparse attack, if the system $(A,C)$ is $2p$-sparse observable, the proposed estimation $\tilde{x}(k)$ solved from (\ref{pb:lasso}) is secure.
\end{theorem}

The proof of Theorem \ref{th:main} is provided in Appendix \ref{ap:th_main} for legibility.
Since sparse observability index only requires simple computation according to Corollary \ref{co:sparse_obs}, this work reduces the complexity of evaluating system vulnerability significantly under the assumption of geometric multiplicity.
For general $A$ that has eigenvalues with geometric multiplicity larger than 1 ($A$ is derogatory), computing sparse observability index is an NP-hard problem \cite{ShoukryTAC2016}\cite{mao2021computational}\cite{yanwen_CDC19}, and there is no computational efficient solution unless P$=$NP. Simultaneously, for algorithm online operation, the computing of our proposed estimation involves solving a convex optimization problem based on LASSO \cite{LASSOTibshirani}, which can be done efficiently.

Moreover, the condition of $2p$-sparse observable is necessary for secure state recovering in the presence of $p$ compromised sensors \cite{ShoukryTAC2016}, which establishes the fundamental limit. Our proposed estimator achieves this limit and also holds optimality in the absence of attack for certain probability, which can be adjusted by parameter $\gamma$, according to Theorem \ref{th:no_attack}. 
By properly tuning parameter $\gamma$, the system operator can achieve the trade-off between the performance in normal operation and performance under attack.

\section{Simulation}\label{sec:sim}
We use an inverted pendulum for the numerical simulation. The physical parameters are illustrated in Fig. \ref{fig:invpen}. 
The control input $u(k)$ is the force applied on the cart, and we assume that there are no frictions of any form.
The states $x_1,x_2,x_3,x_4$ represent cart position coordinate, cart velocity, pendulum angle from vertical and pendulum angle velocity respectively. 
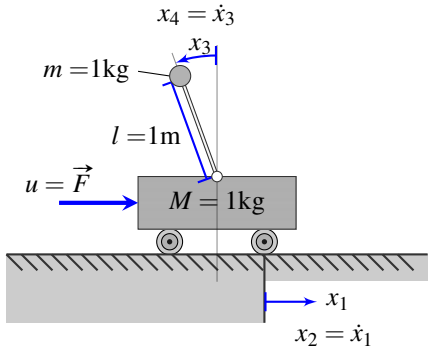
\begin{figure}[htpb]
	\centering\vspace{-10pt}
	\input{inv_pen.tex}\vspace{-10pt}
	\caption{Illustration of the inverted pendulum.}\label{fig:invpen}
\end{figure}

Consider the system linearized at $x_3=x_4=0$, and we sample the continuous-time linear system periodically every $0.1$ seconds. The system matrix $A$ can be transformed into the following Jordan canonical form $$
\begin{bmatrix}
1  & 1 &  0 & 0\\
0  & 1 & 0 &  0\\
0  & 0 & 0.642 & 0  \\
0  & 0 & 0 &  1.557 
\end{bmatrix},
$$ 
and there is a Jordan block with size $2\times 2$ and the geometric multiplicity of all the eigenvalues are 1.

The output matrix $C$ is defined as $C_1=C_2=C_3=[1\ 0\ 0\ 0]$,  $C_4=[0\ 0\ 1\ 0]$, i.e., first three sensors are monitoring the cart position $x_1$ and the fourth sensor is monitoring pendulum angle $x_3$.
In this case, the first 3 sensors can observe all the states, and the fourth sensor can only observe state $x_3$ and $x_4$.
According to Corollary \ref{co:sparse_obs}, the sparse observability index is 2, and the system can at most withstand 1 corrupted sensor.

The noise covariances of the system satisfy $w(k) \sim {N}(0, 0.001\cdot I_4)$ and $v(k) \sim {N}(0, 0.001\cdot I_4)$, where $I_n$ is the identity matrix with size $n\times n$. 
The controller of the system is designed as a Linear-Quadratic Regulator (LQR), and the feedback matrix is chosen as 
$$K_{\rm lqr}=\begin{bmatrix}
-0.604 & -1.678 & -39.514 & -9.721
\end{bmatrix}.
$$
We demonstrate the performance of different estimators of close-loop system where $u(k)=-K_{\rm lqr} x(k)$.

\begin{figure}[htpb]
	\input{states_under_attack_closeloop.tex}
	\caption{Value of system states and their estimation in the presence of attack. The black solid line represents the real state. The red dashed line and blue dotted line represent Kalman estimation and our proposed secure estimation respectively. The initial state is $x(0)=[0,\ 1,\ 0,\ 1]{'}$.} \label{fig:close_loop}
\end{figure}
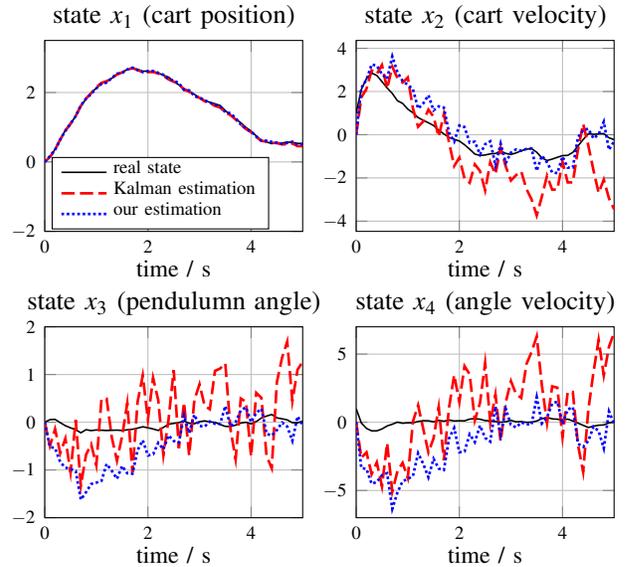

Fig. \ref{fig:close_loop} presents the Kalman estimation and our proposed secure estimation in the presence of attack on sensor 4. The attack $a_4(k)$ is a time-independent random value uniformly distributed on open interval $\left(-\frac{\pi}{2},\frac{\pi}{2}\right)$. In other words, the measurement of the pendulum angle is randomly shifted by the malicious adversary in magnitude of $\pm 90 ^{\circ}$. The attack is launched since the beginning time. As shown in Fig \ref{fig:close_loop}, Kalman estimation (denoted as red dashed line) has large estimation error under the attack while our estimation (denoted as blue dotted line) converges to the real state quickly.

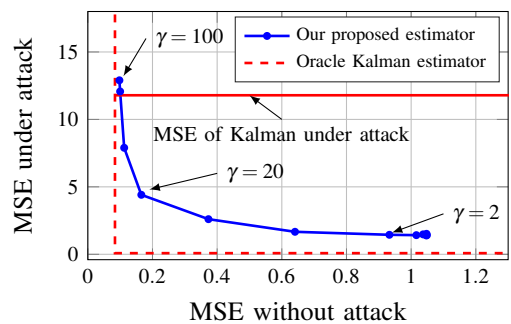
\begin{figure}[htpb]
	\centering
	\input{MSE.tex}
	\caption{Estimation mean square error (MSE) in the absence and in the presence of attack with varying tuning parameter $\gamma$.} \label{fig:MSE}
\end{figure}

Fig. \ref{fig:MSE} illustrates the estimation mean square error (MSE) of our proposed estimator with varying $\gamma$ in the absence and in the presence of attack on sensor 4. 
The attack launched on the system is the same as in Fig. \ref{fig:close_loop}.
The MSE of the oracle Kalman estimation, .i.e., Kalman estimation not affected by the attack, is illustrated by the red dashed line in Fig. \ref{fig:MSE}.
As shown in the figure, by properly choosing $\gamma$, the MSE of our proposed estimator is smaller than that of Kalman estimation (the blue line is below the red horizontal line), with the cost that MSE without attack is slightly larger. 
Moreover, the MSE of our proposed estimator without attack coincides with that of Oracle Kalman filter when $\gamma$ is large (e.g., larger than 50), which validates the performance of our estimator in normal operation. 

\begin{figure}[htpb]
	\centering
	\input{MSE_mag.tex}
	\caption{Estimation mean square error (MSE) in the absence and in the presence of attack with varying attack magnitude $\|a\|_\ift$. The parameter $\gamma$ is set as 5.} \label{fig:MSE_mag}
\end{figure}
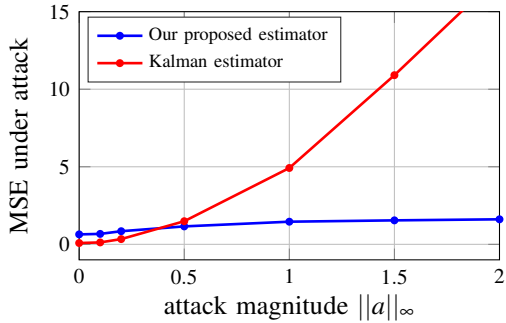

Fig. \ref{fig:MSE_mag} illustrates the estimation mean square error (MSE) of our proposed estimator and Kalman filter with varying attack magnitude $\|a\|_\ift$. The
attack signal $a_4(k)$ is uniformly distributed in interval $(-\|a\|_\ift, \|a\|_\ift)$. As the magnitude of attack signal increases, MSE of Kalman estimator increases significantly while MSE of our proposed secure estimator remains low. This result demonstrates that our proposed estimator holds low estimation error in the presence of malicious bias data in large magnitude. At the same time, in the absence of attack (when $\|a\|_\ift=0$), the estimation error is only slightly larger than that of the optimal Kalman filter.

\section{Conclusion}\label{sec:conclusion}
This paper considers LTI system with Gaussian noise against malicious attack on a subset of sensors. 
We improve upon the previous work by reducing the computational complexity in the designing phase while remaining low computational complexity for online estimation, under the assumption that all the geometric multiplicities of eigenvalues of $A$ are $1$ ($A$ is non-derogatory).
To achieve this, we prove that the span of the rows of $G_i$ is equivalent to the observable space associated to sensor $i$, based on which the canonical form $H_i$ is designed. The proposed estimator is formulated as a convex optimization problem based on $H_i$. We further prove that the proposed estimation is secure if the system is $2p$-sparse observable, which is easily checked due to the simple form of $H_i$. Moreover, in the absence of attack, the proposed estimation coincides with Kalman estimation for certain probability, which can be adjusted by tuning parameter $\gamma$ to balance between the performance with and without attack.


\appendix

\subsection{Proof of Theorem \ref{th:main}}\label{ap:th_main}

Before proving Theorem \ref{th:main}, we need the following Lemma.
Define the number of honest sensors and compromised sensors (w.r.t. compromised set $\Ic$) that can observe state $j$ as:
\begin{align*}
h_j(\Ic)\triangleq |\Sc_j\cap \Ic^c|, \
c_j(\Ic)\triangleq |\Sc_j\cap \Ic|.
\end{align*}
In order to prevent degradation problems, we concentrate on the case where the system sparse observability index is at least $p$. In this case, for any $\Ic$ with $|\Ic|=p$, we have $h_j(\Ic)>c_j(\Ic)\geq 0$.
We have the following lemma quantifying the property of $h_j(\Ic)$ and $c_j(\Ic)$.  
\begin{lemma}\label{lm:cj<hj}
	The following two propositions are equivalent.
	\begin{enumerate}
		\item The system is $2p$-sparse observable.
		\item For any $\Ic$ with $|\Ic|=p$, the inequality $c_j(\Ic)<h_j(\Ic)$ holds for all $j\in\{1,2,\cdots,n\}$.
	\end{enumerate}
\end{lemma}
\begin{proof}(\textbf{Proof of Lemma \ref{lm:cj<hj}})
	We prove the contrapositive of $(1)\Rightarrow(2)$. Supposing that there exists $j^*$ and $\Ic^*$ with $|\Ic^*|=p$ such that $ c_{j^*}(\Ic^*)\geq h_{j^*}(\Ic^*)$, then
	$h_{j^*}(\Ic^*)\leq c_{j^*}(\Ic^*)\leq |\Ic^*|=p$. Noticing that $c_j(\Ic)+h_j(\Ic)=|\Sc_j|$ holds for all $\Ic$, we have $|\Sc_{j^*}|=h_{j^*}(\Ic^*)+c_{j^*}(\Ic^*)\leq 2p$. 
	As a result, there exists set $\Ac$ with $|\Ac|=2p$ such that $\Sc_{j^*}\subseteq\Ac\subseteq\Rc$.
	According to the definition of $\Sc_{j^*}$, there exists no sensor in set $\Rc\setminus \Ac$ who can observe state $j^*$, i.e.,
	\begin{equation*}
	e_{j^*}\notin \rs(O_i),\ \forall i\in\Rc\setminus \Ac.
	\end{equation*}
	As a result, system $(A,C_{\Rc\setminus\Ac})$ is not $2p$-sparse observable according to Definition \ref{df:sparse_obs}. We have completed the proof of $\neg(2)\Rightarrow \neg(1)$, and thus $(1)\Rightarrow(2)$ is proved.
	
	We proceed to prove $(2)\Rightarrow(1)$. 
	We first prove that the system sparse observability index is at least $p$.
	Since if otherwise, there exists a state $j^*$ such that $|\Sc_{j^*}|\leq p$. Choose corrupted set $\Ic^*$ with $p$ elements such that $\Ic^*\supseteq \Sc_{j^*}$. As a result, $h_{j^*}(\Ic^*)=0$ while $c_{j^*}(\Ic^*)=|\Sc_{j^*}|>0$, which contradicts to proposition (2).
	Now we know that the system sparse observability index is at least $p$. Therefore, for each $j\in\{1,2,\cdots,n\}$, there exists an $\Ic^\star$ such that $c_j(\Ic^\star)=p$, and thus $|\Sc_j|=h_j(\Ic^\star)+c_j(\Ic^\star)\geq 2p+1$. According to the definition $\Sc_j$, there are at least $2p+1$ sensors that can observe sensor $j$, and the system is $2p$-sparse observable.	
\end{proof}
	
\begin{proof}(\textbf{Proof of Theorem \ref{th:main}})
	In view of Theorem \ref{th:TAC} and Lemma \ref{lm:cj<hj}, it suffices to prove that the following two propositions are equivalent:
	\begin{enumerate}
		\item The system is $2p$-sparse observable.
		\item The following inequality holds for all $x \neq \mathbf{0}$, $x\in\Rb^n$:
		\begin{equation}\label{eq:proof_cond}
		\sum_{i \in \mathcal{I}}\left\|H_{i} x\right\|_{1}<\sum_{i \in \mathcal{I}^{c}}\left\|H_{i} x\right\|_{1}, \quad \forall\ \Ic\subset \Rc, |\mathcal{I}|\leq p .
		\end{equation}
	\end{enumerate}
	Based on the form of $H_i$ in Theorem \ref{th:span}, inequality \eqref{eq:proof_cond} can be written as 
	\begin{equation*}
	\sum_{j=1}^{n} \sum_{i \in \mathcal{I}\cap\Sc_j}|x_j|<\sum_{j=1}^{n} \sum_{i \in \mathcal{I}^c\cap\Sc_j}|x_j| 
	\end{equation*}
	or equivalently 
	$$
	\sum_{j=1}^{n} c_j(\Ic)\cdot|x_j|<\sum_{j=1}^{n} h_j(\Ic)\cdot|x_j| .
	$$
	If the system is $2p$-sparse observable, we have $c_j(\Ic)<h_j(\Ic)$ for all $j\in\{1,2,\cdots,n\}$. Thus, $\sum_{j=1}^{n} \left(h_j(\Ic)-c_j(\Ic)\right)\cdot|x_j| >0$ holds for all $x\neq \mathbf{0}$.
	If the system is not $2p$-sparse observable, there exists $\Ic^*$ with $|\Ic^*|=p$ and $j^*$ such that $c_{j^*}(\Ic^*)>h_{j^*}(\Ic^*)$. One can design $x=e_{j^*}$, i.e., let only the $j^*$-th entry be 1 and other entries be 0. As a result,
	$$\left(c_{j^*}(\Ic^*)-h_{j^*}(\Ic^*)\right)\cdot|x_{j^*}|>0= \sum_{j\neq j^*}\left(h_{j}(\Ic^*)-c_{j}(\Ic^*)\right)\cdot|x_j| .$$
	Therefore, condition \eqref{eq:proof_cond} is violated and the proof is completed. 
\end{proof}

\balance



\bibliographystyle{IEEEtran}
\bibliography{ref_zishuo}

\addtolength{\textheight}{-12cm}   

\end{document}

%% file: inv_pen.tex
\begin{tikzpicture}[> = latex',%
	scale = 0.7]
	\tikzset{%
		interface/.style={
			postaction={draw, decorate, decoration={border, angle=-45,
					amplitude=0.3cm, segment length=2mm}}},
		helparrow/.style={>=latex', blue, thick},
		helpline/.style={thin, black!90, opacity=0.5},
		force/.style={>=stealth, draw=blue, fill=blue, ultra thick},
	}
	\def\ground{%
		\fill [black!20] (0, 0) rectangle (49mm, -13mm);
		\fill [black!20] (49mm, 0) rectangle (80mm, -5mm);
		\draw [thick, black!80, interface] (0, 0) -- (80mm, 0);
		\draw [thick, black!80] (49mm, 0) -- +(0, -13mm);
		\draw [|->, helparrow] (49mm, -9mm) -- ++(10mm, 0) node [right] {\color{black} $x_1$} node [below=5pt] {\color{black}  \hspace{10pt} $x_2=\dot{x}_1$};
	}
	
	\def\cart{
		\filldraw [
		draw = black!80,%
		fill = black!40,%
		top color = black!30,%
		bottom color = black!30,%
		] (0,0) rectangle (30mm, 10mm);
		\node at (15mm, 5mm) {$M=1$kg};
		\draw[->, force] (-15mm, 5mm) node [above] {$u=\ensuremath{\vv{F}}$} -- (0, 5mm);
	}
	
	\def\pendulum{%
		\filldraw [
		draw = black!80,%
		left color=black!0,%
		bottom color=black!0,%
		] (-0.3mm, 0) rectangle (0.3mm, 20mm);
		\draw [|-|, helparrow] (-2mm, 0mm) -- ++(0mm, 20mm);
		\node at (-10mm, 12mm) {$l=$1m};
	}
	
	\def\joint{%
		\filldraw [
		draw = black!80,%
		fill = white,%
		] (0, 0) circle (1mm);
	}
\def\ball{%
	\filldraw [
	draw = black!80,%
	fill = black!30,%
	] (0, 0) circle (2mm);
}
	
	\def\wheel{%
		\fill [thin,%
		fill = black!30,%
		] (0, 0) circle (1.75mm);
		\begin{scope}
			\clip (0, 0) circle (1.75mm);
			\fill [fill=black!30,] (0, -1mm) circle (2mm);
		\end{scope}
		\fill [fill=black!90] (0, 0) circle (0.5mm);
		\draw [thin,%
		double = black!20,%
		double distance = 0.5mm] (0, 0) circle (2mm);
	}

	\begin{scope}
		\wheel
	\end{scope}
	\begin{scope}[xshift=18mm]
		\wheel
	\end{scope}
	\begin{scope}[xshift=-31mm, yshift=-2.4mm]
		\ground
	\end{scope}
	\begin{scope}[shift = {(-6mm, 2.4mm)}]
		\cart
	\end{scope}
	\begin{scope}[shift = {(9mm, 12.4mm)}]
		\draw [helpline] (0, -20mm) -- (0, 25mm);
		\draw [helpline] (110:20.5mm) -- (110:25mm);
		\draw [|->, helparrow] (0, 23mm) 
		arc [radius=23mm, start angle=90, delta angle=20] ;
		\node at (97:25mm) {$x_3$}; \node at (97:31mm) {$x_4=\dot{x}_3$};
		\draw [thin] (110:20mm) -- (-15mm, 20mm) 
		node [left] {$m=$1kg};
	\end{scope}
	\begin{scope}[shift = {(9mm, 12.4mm)}, rotate=20]
		\pendulum
	\end{scope}
	\begin{scope}[shift = {(9mm, 12.4mm)}]
		\joint
	\end{scope}
	\begin{scope}[shift = {(2mm, 31.5mm)}]
		\ball
	\end{scope}
\end{tikzpicture}


%% file: states_under_attack_closeloop.tex
\begin{tikzpicture}
\begin{axis}
[	width=1.35in,
	height=1.0in,
	at={(0in,1.5in)},
	scale only axis,
	xmin=0,
	xmax=5.0,
	ymin=-2,
	ymax=3.5,
	xlabel={\small time / s},
	title={state $x_1$ (cart position)},
	title style={at={(axis description cs:0.5,1.15)},anchor=north},
	x label style={at={(axis description cs:0.5,-0.1)},anchor=north},
	y label style={at={(axis description cs:-0.11,.5)},anchor=south},
	xticklabel style = {font=\scriptsize},
	yticklabel style = {font=\scriptsize},
	axis background/.style={fill=white},
	xmajorgrids,
	ymajorgrids,
	legend style={at={(0.87,0.01)}, anchor=south east, legend cell align=left, align=left, text height=0.4ex,text depth=1.1ex,row sep=0.02cm, font=\scriptsize}]
	 \addplot [color={black}, line width=0.6pt]
	table[row sep={\\}]
	{ 0.0  0.0  \\
		0.1  0.13057178855316634  \\
		0.2  0.3669336773273081  \\
		0.3  0.6393037558605023  \\
		0.4  0.9139987980819133  \\
		0.5  1.0807907594390098  \\
		0.6  1.3358683761582224  \\
		0.7  1.598132092764893  \\
		0.8  1.8210645726884904  \\
		0.9  1.992870298275197  \\
		1.0  2.1248543704609557  \\
		1.1  2.2590955089417584  \\
		1.2  2.312867406060577  \\
		1.3  2.4119274641783375  \\
		1.4  2.5314430122083493  \\
		1.5  2.58432224039452  \\
		1.6  2.6625567947648787  \\
		1.7  2.681731830734295  \\
		1.8  2.667272156664494  \\
		1.9  2.6159063964191214  \\
		2.0  2.6143615456342038  \\
		2.1  2.6059644956834367  \\
		2.2  2.5605932075102  \\
		2.3  2.4976198188561933  \\
		2.4  2.442767646770862  \\
		2.5  2.3177815258335093  \\
		2.6  2.2207158386304684  \\
		2.7  2.1105646038953436  \\
		2.8  2.0426455375695327  \\
		2.9  1.9476042715764184  \\
		3.0  1.8508977394534254  \\
		3.1  1.7701062182874918  \\
		3.2  1.717474598746494  \\
		3.3  1.6744632402613224  \\
		3.4  1.6240091830535683  \\
		3.5  1.4791317160530866  \\
		3.6  1.3492186238106822  \\
		3.7  1.2368051937883977  \\
		3.8  1.1303811993177615  \\
		3.9  0.9925879266051355  \\
		4.0  0.8908101438356706  \\
		4.1  0.7496274213992153  \\
		4.2  0.617917318074613  \\
		4.3  0.5757141323689275  \\
		4.4  0.5273881701167311  \\
		4.5  0.5539046402372925  \\
		4.6  0.5438774346162495  \\
		4.7  0.5727580160330278  \\
		4.8  0.5644188181116768  \\
		4.9  0.5349149205017439  \\
		5.0  0.5300540948931045  \\
	}
	;\addlegendentry{real state};
	
	\addplot [color={red},dash pattern={on 5pt off 2pt}, line width=1pt]
	table[row sep={\\}]
	{     0.0  0.0  \\
		0.1  0.14552418478604176  \\
		0.2  0.34997892894398636  \\
		0.3  0.6465143081042575  \\
		0.4  0.9024646402843046  \\
		0.5  1.1026710582693795  \\
		0.6  1.3207084438311545  \\
		0.7  1.6166585089311911  \\
		0.8  1.824408537639195  \\
		0.9  1.9822890683059842  \\
		1.0  2.1549982416313753  \\
		1.1  2.277567629811386  \\
		1.2  2.327690462107993  \\
		1.3  2.4096348063297235  \\
		1.4  2.5204859943199236  \\
		1.5  2.5776035392624834  \\
		1.6  2.6618596935446517  \\
		1.7  2.707143325993244  \\
		1.8  2.651316455906345  \\
		1.9  2.5851842920662604  \\
		2.0  2.5879299056886262  \\
		2.1  2.5919757876528977  \\
		2.2  2.5469132831404013  \\
		2.3  2.4852446952159806  \\
		2.4  2.378984306403647  \\
		2.5  2.310765395367408  \\
		2.6  2.2441624056891216  \\
		2.7  2.080062036769485  \\
		2.8  2.0261270807231737  \\
		2.9  1.8969070317854655  \\
		3.0  1.847154026663423  \\
		3.1  1.747844041250124  \\
		3.2  1.6927658216903954  \\
		3.3  1.6157147934986855  \\
		3.4  1.568168522120093  \\
		3.5  1.42293344838227  \\
		3.6  1.3487689446426436  \\
		3.7  1.2242109161732482  \\
		3.8  1.0781475744380853  \\
		3.9  0.9490134907742411  \\
		4.0  0.8571653746196196  \\
		4.1  0.7192584722957089  \\
		4.2  0.6088006197850202  \\
		4.3  0.5813735203757455  \\
		4.4  0.5534343607684401  \\
		4.5  0.5415339957210873  \\
		4.6  0.5056362337476169  \\
		4.7  0.5085720520871548  \\
		4.8  0.5307678302538518  \\
		4.9  0.4565711456370377  \\
		5.0  0.45278335598162417  \\
	}
	;\addlegendentry{Kalman estimation};
	
    \addplot[densely dotted, color={blue}, line width=1pt]
        table[row sep={\\}]
        {   
        	0.0  0.0  \\
        	0.1  0.1657851657197167  \\
        	0.2  0.35079167965465663  \\
        	0.3  0.645776596773159  \\
        	0.4  0.8979336688741562  \\
        	0.5  1.1037334484021815  \\
        	0.6  1.3164173608504213  \\
        	0.7  1.630384029215857  \\
        	0.8  1.8193184367673567  \\
        	0.9  1.9581751546009887  \\
        	1.0  2.120984114948585  \\
        	1.1  2.2506654853807713  \\
        	1.2  2.2862641979189453  \\
        	1.3  2.4197002565701125  \\
        	1.4  2.5426140750756057  \\
        	1.5  2.5609169458577945  \\
        	1.6  2.665558426914958  \\
        	1.7  2.7148107442178953  \\
        	1.8  2.644876489580162  \\
        	1.9  2.5883422186087706  \\
        	2.0  2.616091169386526  \\
        	2.1  2.620269523821148  \\
        	2.2  2.569287779885978  \\
        	2.3  2.5097541624556  \\
        	2.4  2.3896098913727086  \\
        	2.5  2.3436346314070087  \\
        	2.6  2.272193433534641  \\
        	2.7  2.0754237876540453  \\
        	2.8  2.061443572443403  \\
        	2.9  1.9017672251886233  \\
        	3.0  1.8568362057040122  \\
        	3.1  1.7760806970540275  \\
        	3.2  1.7090671070157608  \\
        	3.3  1.6436635268720843  \\
        	3.4  1.598098019821304  \\
        	3.5  1.4332853185642722  \\
        	3.6  1.4031459280319951  \\
        	3.7  1.2612505774097464  \\
        	3.8  1.0806601187472133  \\
        	3.9  0.9620691052927659  \\
        	4.0  0.86786843433087  \\
        	4.1  0.7368596571623817  \\
        	4.2  0.6080403846652683  \\
        	4.3  0.6104518033003847  \\
        	4.4  0.5770089682008476  \\
        	4.5  0.550992751696677  \\
        	4.6  0.5192274796280982  \\
        	4.7  0.5448226204251811  \\
        	4.8  0.5821849829819542  \\
        	4.9  0.4997697749898838  \\
        	5.0  0.5109474057098409  \\
        }
        ;\addlegendentry{our estimation};
   
\end{axis}

\begin{axis}
[	width=1.35in,
height=1.0in,
at={(1.63in,1.5in)},
scale only axis,
xmin=0,
xmax=5.0,
xlabel={\small time / s},
title={state $x_2$ (cart velocity)},
title style={at={(axis description cs:0.5,1.15)},anchor=north},
x label style={at={(axis description cs:0.5,-0.1)},anchor=north},
y label style={at={(axis description cs:-0.11,.5)},anchor=south},
xticklabel style = {font=\scriptsize},
yticklabel style = {font=\scriptsize},
axis background/.style={fill=white},
xmajorgrids,
ymajorgrids,
legend style={at={(0.98,0.02)}, anchor=south east, legend cell align=left, align=left, draw=white!15!black, font=\scriptsize}]
\addplot [color={black}, line width=0.6pt]
table[row sep={\\}]
{
    0.0  1.0  \\
   0.1  2.09835633692394  \\
   0.2  2.596534428155504  \\
   0.3  2.852236195369671  \\
   0.4  2.7422797599859483  \\
   0.5  2.4575122604242843  \\
   0.6  2.165388599468268  \\
   0.7  1.9097693406326162  \\
   0.8  1.5719134087785709  \\
   0.9  1.4492075318464543  \\
   1.0  1.184475689586946  \\
   1.1  1.0149497430428243  \\
   1.2  0.7958530642123849  \\
   1.3  0.6649480881334213  \\
   1.4  0.5526175057761921  \\
   1.5  0.42856328102364105  \\
   1.6  0.20671770123581357  \\
   1.7  0.03677065030369919  \\
   1.8  -0.125797153396115  \\
   1.9  -0.21077920181676005  \\
   2.0  -0.29326986034870606  \\
   2.1  -0.4626939314870252  \\
   2.2  -0.6692588306858648  \\
   2.3  -0.8941473732042691  \\
   2.4  -0.9623675545639064  \\
   2.5  -0.9377712739740274  \\
   2.6  -0.886668113213611  \\
   2.7  -0.8700179648481904  \\
   2.8  -0.75232798100226  \\
   2.9  -0.8636187117053217  \\
   3.0  -0.9150779785350952  \\
   3.1  -0.8429512915631019  \\
   3.2  -0.7431352603146522  \\
   3.3  -0.6701500890430048  \\
   3.4  -0.6460644250116183  \\
   3.5  -0.8107806343554698  \\
   3.6  -1.0722315232669763  \\
   3.7  -1.1948854467784327  \\
   3.8  -1.1286114753648142  \\
   3.9  -1.046153331452478  \\
   4.0  -0.966695299452097  \\
   4.1  -0.9562828380902139  \\
   4.2  -0.8175866544263293  \\
   4.3  -0.5173283911342638  \\
   4.4  -0.22415901695197327  \\
   4.5  0.04741282233635041  \\
   4.6  0.016564074160940218  \\
   4.7  0.034991452280959456  \\
   4.8  0.050398480208050406  \\
   4.9  -0.14181871253961592  \\
   5.0  -0.22961616720338732  \\
}
;

\addplot [color={red},dash pattern={on 5pt off 2pt}, line width=1pt]
table[row sep={\\}]
{
     0.0  0.0  \\
   0.1  1.7240777886203358  \\
   0.2  2.1271839540756745  \\
   0.3  2.9783059092821142  \\
   0.4  2.808738759386001  \\
   0.5  3.224386742178278  \\
   0.6  2.1331872927773268  \\
   0.7  3.1489703433395135  \\
   0.8  2.620091938101374  \\
   0.9  2.4149645350954514  \\
   1.0  2.6474500666839123  \\
   1.1  1.219496592945067  \\
   1.2  0.36718814550277745  \\
   1.3  0.6156409868653556  \\
   1.4  1.1619278001323634  \\
   1.5  0.8314257794411613  \\
   1.6  -0.10082962334020129  \\
   1.7  0.9574483000540961  \\
   1.8  -0.3639009640341542  \\
   1.9  -1.448291965209572  \\
   2.0  -1.0623077901480324  \\
   2.1  -2.0259666778950973  \\
   2.2  -2.2245091279548825  \\
   2.3  -1.3707089469977793  \\
   2.4  -1.6769054889875017  \\
   2.5  -2.5741849959924012  \\
   2.6  -1.2794340651027918  \\
   2.7  -1.8269210159647145  \\
   2.8  -0.6024862818459553  \\
   2.9  -1.4739618132321526  \\
   3.0  -2.1196122363772445  \\
   3.1  -1.8705362809211492  \\
   3.2  -1.7031687706780028  \\
   3.3  -2.510468184324549  \\
   3.4  -2.992268440609813  \\
   3.5  -3.7361641464515656  \\
   3.6  -2.8608933000508068  \\
   3.7  -1.895977986525779  \\
   3.8  -2.543967527688811  \\
   3.9  -2.531419778248815  \\
   4.0  -1.2526533231797266  \\
   4.1  -2.2071643322223893  \\
   4.2  -2.207218663845549  \\
   4.3  -0.6727103668042755  \\
   4.4  0.5012765948518413  \\
   4.5  -0.5000257307237576  \\
   4.6  -1.6491234466231006  \\
   4.7  -2.690330083761561  \\
   4.8  -2.077917276366079  \\
   4.9  -2.9392520637772876  \\
   5.0  -3.453293548671936  \\
}
;

\addplot[densely dotted, color={blue}, line width=1pt]
table[row sep={\\}]
{
         0.0  0.0  \\
        0.1  2.164353842656242  \\
        0.2  2.54373251976346  \\
        0.3  3.2528954039442897  \\
        0.4  3.192055234025908  \\
        0.5  2.799412641447088  \\
        0.6  2.5850605432819562  \\
        0.7  3.6259786688571705  \\
        0.8  2.8920491398535995  \\
        0.9  2.488222684858382  \\
        1.0  2.30705657624711  \\
        1.1  2.108723788554722  \\
        1.2  1.207800442546198  \\
        1.3  1.4483169169762302  \\
        1.4  1.8300736063216998  \\
        1.5  1.1274406002311421  \\
        1.6  1.3136305951104776  \\
        1.7  1.2567058910627724  \\
        1.8  0.0915631438922776  \\
        1.9  -0.3180046677866736  \\
        2.0  0.4325637373781553  \\
        2.1  0.3571408235167533  \\
        2.2  -0.3720070787657012  \\
        2.3  -0.32354788586892624  \\
        2.4  -0.9113715080841805  \\
        2.5  -0.439853879505826  \\
        2.6  -0.30375334291964734  \\
        2.7  -1.4274388177664705  \\
        2.8  -0.4137338518116477  \\
        2.9  -1.3066114069577461  \\
        3.0  -0.8257458442321851  \\
        3.1  -0.6789526308080525  \\
        3.2  -0.7464124368554138  \\
        3.3  -0.656193771613524  \\
        3.4  -0.640360456944085  \\
        3.5  -1.6711584409475662  \\
        3.6  -0.9674577980085555  \\
        3.7  -1.2819301317132612  \\
        3.8  -1.7433397028694553  \\
        3.9  -1.7768336031088092  \\
        4.0  -1.011946538964856  \\
        4.1  -1.5230533396036912  \\
        4.2  -1.5175045035977768  \\
        4.3  -0.11861543151092767  \\
        4.4  0.20774606161971773  \\
        4.5  0.27724236220344295  \\
        4.6  -0.17259053586140277  \\
        4.7  0.015788683757612373  \\
        4.8  0.5979215720906388  \\
        4.9  -0.6989243442880786  \\
        5.0  -0.3044707453418079  \\
}
;

\end{axis}

 \begin{axis}
	[	width=1.35in,
	height=1.0in,
	at={(0in,0in)},
	scale only axis,
	xmin=0,
	xmax=5.0,
	ymin=-2,
	ymax=2,
	xlabel={\small time / s},
	title={state $x_3$ (pendulumn angle)},
	title style={at={(axis description cs:0.5,1.15)},anchor=north},
	x label style={at={(axis description cs:0.5,-0.1)},anchor=north},
	y label style={at={(axis description cs:-0.11,.5)},anchor=south},
	xticklabel style = {font=\scriptsize},
	yticklabel style = {font=\scriptsize},
	axis background/.style={fill=white},
	xmajorgrids,
	ymajorgrids,]
	\addplot [color={black}, line width=0.6pt]
	table[row sep={\\}]
	{ 0.0  0.0  \\
		0.1  0.05441543956867995  \\
		0.2  0.06093118739834458  \\
		0.3  -0.00821463634676167  \\
		0.4  -0.07213529373093339  \\
		0.5  -0.1166990412524242  \\
		0.6  -0.1568427403970221  \\
		0.7  -0.22101528895059347  \\
		0.8  -0.15563838503934882  \\
		0.9  -0.17452660522540422  \\
		1.0  -0.16086426600531087  \\
		1.1  -0.16651229616473343  \\
		1.2  -0.17231474030220228  \\
		1.3  -0.1678372289192149  \\
		1.4  -0.16413428893967696  \\
		1.5  -0.17202930514304626  \\
		1.6  -0.15839720137968225  \\
		1.7  -0.14664893061805612  \\
		1.8  -0.12327344555377336  \\
		1.9  -0.09379551278191019  \\
		2.0  -0.11516491358299302  \\
		2.1  -0.14007847985686955  \\
		2.2  -0.17182042264257424  \\
		2.3  -0.10788926904228839  \\
		2.4  -0.040928780511055994  \\
		2.5  -0.023389225499430066  \\
		2.6  -0.007709893819070413  \\
		2.7  0.028345786430855446  \\
		2.8  -0.021281578558225683  \\
		2.9  -0.03186318116009481  \\
		3.0  0.012104461975826902  \\
		3.1  0.02968981582922832  \\
		3.2  0.008232957023269188  \\
		3.3  -0.0006756120685962367  \\
		3.4  -0.04825482302417893  \\
		3.5  -0.08587406992640459  \\
		3.6  -0.08962153323429131  \\
		3.7  -0.058960073581083364  \\
		3.8  -0.01596160363780943  \\
		3.9  -0.00013856303792324784  \\
		4.0  -0.03391915627132086  \\
		4.1  0.0133254995006451  \\
		4.2  0.09028087252500197  \\
		4.3  0.1296859125239912  \\
		4.4  0.1631364003527701  \\
		4.5  0.09197748055248708  \\
		4.6  0.0723380969198081  \\
		4.7  0.05001548420747059  \\
		4.8  -0.008428051574268116  \\
		4.9  -0.01228837292483571  \\
		5.0  0.024936944827092425  \\
	}
	;
	\addplot [color={red},dash pattern={on 5pt off 2pt}, line width=1pt]
	table[row sep={\\}]
	{ 0.0  0.0  \\
		0.1  -0.4958208513293527  \\
		0.2  -0.14244600424469694  \\
		0.3  -0.64822091941416  \\
		0.4  -0.34091145038085785  \\
		0.5  -1.0488891950124752  \\
		0.6  0.046840195431606446  \\
		0.7  -1.3257379017272237  \\
		0.8  -0.6822363630132716  \\
		0.9  -0.4609216892083121  \\
		1.0  -0.8989636678046296  \\
		1.1  0.4878914173424605  \\
		1.2  0.4926821180434989  \\
		1.3  -0.31932718274193506  \\
		1.4  -0.8037849342093336  \\
		1.5  -0.29161233265293746  \\
		1.6  0.3991645618923523  \\
		1.7  -1.0513629779999338  \\
		1.8  0.42947779594085317  \\
		1.9  0.980258501749623  \\
		2.0  0.058139200217865616  \\
		2.1  0.9298336360414171  \\
		2.2  0.46445516287531485  \\
		2.3  -0.6886361666344246  \\
		2.4  0.06032649512994978  \\
		2.5  1.0923158132084432  \\
		2.6  -0.540518093210258  \\
		2.7  0.41604776413397077  \\
		2.8  -0.6948728469862149  \\
		2.9  0.4633142764726519  \\
		3.0  0.7974437778872434  \\
		3.1  0.34217812501061706  \\
		3.2  0.27261180841217103  \\
		3.3  1.077878631870225  \\
		3.4  1.1262176810628386  \\
		3.5  1.234170892867463  \\
		3.6  -0.144442039067605  \\
		3.7  -0.771822190051634  \\
		3.8  0.4985922594982327  \\
		3.9  0.4287344163992547  \\
		4.0  -0.7152409698831131  \\
		4.1  0.6161203516934782  \\
		4.2  0.5127453069105411  \\
		4.3  -0.7097679812442066  \\
		4.4  -0.9624561752584151  \\
		4.5  0.8431230001015212  \\
		4.6  1.3524455501329735  \\
		4.7  1.6877686591841392  \\
		4.8  0.4469811895430389  \\
		4.9  1.0988304482327425  \\
		5.0  1.2729975819743555  \\
	}
	;
	
	\addplot[densely dotted, color={blue}, line width=1pt]
	table[row sep={\\}]
	{
    0.0  0.0  \\
 0.1  -0.514510861498952  \\
 0.2  -0.5428580728632053  \\
 0.3  -0.8350876716599206  \\
 0.4  -0.9498686148298572  \\
 0.5  -0.9790108195246354  \\
 0.6  -1.0212271492894445  \\
 0.7  -1.6214907359471304  \\
 0.8  -1.4647943037325368  \\
 0.9  -1.292686065367979  \\
 1.0  -1.2656252227293743  \\
 1.1  -1.2274459868268626  \\
 1.2  -0.8726269997615367  \\
 1.3  -1.0160739558463279  \\
 1.4  -1.1864024116249163  \\
 1.5  -0.9137548968310937  \\
 1.6  -1.0695279196732854  \\
 1.7  -1.078652929436544  \\
 1.8  -0.5852943936384741  \\
 1.9  -0.37676250726902016  \\
 2.0  -0.6947083621183976  \\
 2.1  -0.6708755347186818  \\
 2.2  -0.4321199405881123  \\
 2.3  -0.5662464607138092  \\
 2.4  -0.2756346696974078  \\
 2.5  -0.39210156744508623  \\
 2.6  -0.42667301613082065  \\
 2.7  0.14602886191584957  \\
 2.8  -0.22573177308226128  \\
 2.9  0.15977717674625644  \\
 3.0  -0.07795775516620629  \\
 3.1  -0.09139370324042077  \\
 3.2  0.017368953305363824  \\
 3.3  -0.03159010629161228  \\
 3.4  -0.03365924153443705  \\
 3.5  0.34083355862784487  \\
 3.6  -0.08633502220156827  \\
 3.7  0.0047686052904942905  \\
 3.8  0.24843534607612908  \\
 3.9  0.31480373409980716  \\
 4.0  9.625258913274412e-5  \\
 4.1  0.19586522575290582  \\
 4.2  0.2851722272514943  \\
 4.3  -0.239779072681364  \\
 4.4  -0.28004425594484  \\
 4.5  -0.16343708613672764  \\
 4.6  -0.006012839202030086  \\
 4.7  -0.08838550658112805  \\
 4.8  -0.38894356668291413  \\
 4.9  0.12195221980706963  \\
 5.0  -0.0617899124381496  \\
	}
;
\end{axis}

\begin{axis}
[	width=1.35in,
height=1.0in,
at={(1.63in,0in)},
scale only axis,
xmin=0,
xmax=5.0,
ymin=-7,
ymax=7,
xlabel={\small time / s},
title={state $x_4$ (angle velocity)},
title style={at={(axis description cs:0.5,1.15)},anchor=north},
x label style={at={(axis description cs:0.5,-0.1)},anchor=north},
y label style={at={(axis description cs:-0.11,.5)},anchor=south},
xticklabel style = {font=\scriptsize},
yticklabel style = {font=\scriptsize},
axis background/.style={fill=white},
xmajorgrids,
ymajorgrids,]
\addplot [color={black}, line width=0.6pt]
table[row sep={\\}]
{
     0.0  1.0  \\
     0.1  -0.049774922644450825  \\
     0.2  -0.453551724665157  \\
     0.3  -0.6289877263039896  \\
     0.4  -0.6302452819309194  \\
     0.5  -0.4911000412108106  \\
     0.6  -0.26994766723324126  \\
     0.7  -0.1444437334465279  \\
     0.8  0.009809636998514892  \\
     0.9  -0.07196680190062843  \\
     1.0  -0.03840375977024918  \\
     1.1  -0.03916075975803085  \\
     1.2  0.056437291344504203  \\
     1.3  0.11687575372767825  \\
     1.4  0.10919845333166542  \\
     1.5  0.030710271651288867  \\
     1.6  0.14040855006287534  \\
     1.7  0.07760427426811214  \\
     1.8  0.16314424278586556  \\
     1.9  0.09631485919212823  \\
     2.0  0.11032192093448426  \\
     2.1  0.10372985385087166  \\
     2.2  0.23178961726544062  \\
     2.3  0.27140793710551414  \\
     2.4  0.22399508371035967  \\
     2.5  0.12524696840908248  \\
     2.6  0.10028575714171252  \\
     2.7  0.05255407502335051  \\
     2.8  0.006053264399104558  \\
     2.9  0.03714061638900377  \\
     3.0  0.14073602295764176  \\
     3.1  0.0320471643671533  \\
     3.2  0.049808476430819607  \\
     3.3  -0.0036373982636823746  \\
     3.4  0.012469467726322462  \\
     3.5  0.09219161836191604  \\
     3.6  0.26690372801874823  \\
     3.7  0.35507336551527247  \\
     3.8  0.2505743916264524  \\
     3.9  0.224104385151202  \\
     4.0  0.16152298633293932  \\
     4.1  0.23386272213617237  \\
     4.2  0.09735885168048589  \\
     4.3  -0.06346520347028528  \\
     4.4  -0.2246304754727138  \\
     4.5  -0.3810429327988814  \\
     4.6  -0.28813439896871107  \\
     4.7  -0.21694244708778623  \\
     4.8  -0.1941849312479038  \\
     4.9  -0.0470572718355019  \\
     5.0  0.009424920046830293  \\
}
;
\addplot [color={red},dash pattern={on 5pt off 2pt}, line width=1pt]
table[row sep={\\}]
{
     0.0  0.0  \\
    0.1  -2.602714374173333  \\
    0.2  -2.206137327684231  \\
    0.3  -3.8019680786359475  \\
    0.4  -2.941134056692553  \\
    0.5  -4.701901447373908  \\
    0.6  -1.5213760429104186  \\
    0.7  -5.08296582525162  \\
    0.8  -3.5841673883540475  \\
    0.9  -2.973702297373957  \\
    1.0  -4.078053104058243  \\
    1.1  0.05252387232433153  \\
    1.2  1.24779831673841  \\
    1.3  -0.47937762186832966  \\
    1.4  -2.2102813452879775  \\
    1.5  -1.1735395266307695  \\
    1.6  1.121512470804875  \\
    1.7  -2.7923828384503997  \\
    1.8  1.0840097698830888  \\
    1.9  3.4663736867151957  \\
    2.0  1.4866756603702052  \\
    2.1  4.119471883144275  \\
    2.2  3.5184664150469764  \\
    2.3  0.1437445292685524  \\
    2.4  1.4192557765574065  \\
    2.5  4.342686930741751  \\
    2.6  -0.04030506460730354  \\
    2.7  2.035931851129575  \\
    2.8  -1.404102962326088  \\
    2.9  1.4838250114228737  \\
    3.0  3.0180069991155216  \\
    3.1  2.0653980580679168  \\
    3.2  1.7836228580829523  \\
    3.3  4.259822088670521  \\
    3.4  5.1600164056166475  \\
    3.5  6.299599126009034  \\
    3.6  2.707456619902828  \\
    3.7  -0.04536359524649569  \\
    3.8  2.635325595030401  \\
    3.9  2.5374419451437955  \\
    4.0  -1.0809958289324628  \\
    4.1  2.2131461854344656  \\
    4.2  2.222979256831081  \\
    4.3  -1.6592604189460227  \\
    4.4  -3.631901224835614  \\
    4.5  0.9101383775164766  \\
    4.6  3.683566027444797  \\
    4.7  6.0468347688540645  \\
    4.8  3.4388530869384875  \\
    4.9  5.53268690798357  \\
    5.0  6.582179675696852  \\
}
;

\addplot[densely dotted, color={blue}, line width=1pt]
table[row sep={\\}]
{
     0.0  0.0  \\
   0.1  -3.3562468838241544  \\
   0.2  -3.421097619583297  \\
   0.3  -4.51004020327534  \\
   0.4  -4.379462554860902  \\
   0.5  -3.884711639835924  \\
   0.6  -3.7354876527447414  \\
   0.7  -6.346769138745568  \\
   0.8  -5.148645980660467  \\
   0.9  -4.322601624906548  \\
   1.0  -4.161372545936952  \\
   1.1  -3.939524456185833  \\
   1.2  -2.262527384182802  \\
   1.3  -3.045454761059592  \\
   1.4  -4.036462511829534  \\
   1.5  -2.6794118555479867  \\
   1.6  -3.3984646871356445  \\
   1.7  -3.4976896964599664  \\
   1.8  -1.1826754282508036  \\
   1.9  -0.4192552103947449  \\
   2.0  -2.198719332991274  \\
   2.1  -2.142721367641647  \\
   2.2  -0.8782340890798431  \\
   2.3  -1.45311690137029  \\
   2.4  -0.24072722888125228  \\
   2.5  -1.1393848484304567  \\
   2.6  -1.4162623767059277  \\
   2.7  1.1521534240398912  \\
   2.8  -0.9108602736186713  \\
   2.9  0.9739239773894761  \\
   3.0  -0.196922032651274  \\
   3.1  -0.374403189778348  \\
   3.2  -0.01235238789516525  \\
   3.3  -0.14084493785971225  \\
   3.4  -0.1627997474906543  \\
   3.5  1.8381478993591889  \\
   3.6  -0.08738267688267179  \\
   3.7  0.3631260471174729  \\
   3.8  1.3637068919434006  \\
   3.9  1.517266885126714  \\
   4.0  -0.1321363930829356  \\
   4.1  0.9365976888908358  \\
   4.2  1.1101596209148719  \\
   4.3  -1.5790809439202156  \\
   4.4  -1.8026772144271552  \\
   4.5  -1.371786649528542  \\
   4.6  -0.3256850694406754  \\
   4.7  -0.6435369243065475  \\
   4.8  -1.907503673473588  \\
   4.9  0.7980769770961511  \\
   5.0  -0.1285047621574341  \\
}
;

\end{axis}

\end{tikzpicture}

%% file: MSE.tex
\begin{tikzpicture}

\begin{axis}
	[width=2.2in,
	height=1.3in,
	at={(0in,1.5in)},
	scale only axis,
	xmin=0,
	xmax=1.3,
	ymin=-0.4,
	ymax=18,
	xlabel={MSE without attack},
	ylabel={MSE under attack},
	y label style={at={(axis description cs:-0.11,.5)},anchor=south},
	xticklabel style = {font=\footnotesize},
	yticklabel style = {font=\footnotesize},
	axis background/.style={fill=white},
	xmajorgrids,
	ymajorgrids,
	legend style={at={(0.98,0.98)}, anchor=north east, legend cell align=left, align=left, draw=white!15!black, font=\scriptsize}]

		\addplot [color={blue}, line width=1pt, mark=*, mark size=1pt]
		table[row sep={\\}]
		{   
	 1.04751 1.51703\\
	1.04864 1.41575 \\
	1.04684 1.39841 \\
	1.03689 1.47677\\
	1.01606 1.4194 \\
	0.932737  1.44531 \\
	0.640747 1.66763\\
	0.37311 2.60405\\
	0.16548  	4.40994\\
	0.11245   7.895264\\
	0.0999183	12.0712\\
 	0.0976583 	12.8994\\
		}
		;\addlegendentry{Our proposed estimator}

	\addplot [dashed, color={red}, line width=1pt]
	table[row sep={\\}]
	{
		3   0.0839695  \\
		 0.0839695   0.0839695\\
		 0.0839695  40\\
	}
	;\addlegendentry{Oracle Kalman estimator}
	
	 \addplot [color={red}, line width=1pt]
	table[row sep={\\}]
	{
		 0.0839695  11.78984562427724\\
		3  11.78984562427724\\
	}
	;

	\draw[latex-] (0.11,13.2)--(0.18,16) node[right]{\footnotesize $\gamma=100$};
	\draw[latex-] (0.19,4.8)--(0.4,6) node[right]{\footnotesize $\gamma=20$};
	\draw[latex-] (0.94,1.6)--(1.1,3) node[right]{\footnotesize $\gamma=2$};
	\draw[latex-] (0.5, 11.78)--(0.6,10) node[below=-1pt]{\footnotesize MSE of Kalman under attack};
	
\end{axis}

\end{tikzpicture}

%% file: MSE_mag.tex
\begin{tikzpicture}
\begin{axis}
	[width=2.2in,
	height=1.3in,
	at={(3.5in,0in)},
	scale only axis,
	xmin=0,
	xmax=2.0,
	ymin=-1,
	ymax=15,
	xlabel={attack magnitude $||a||_\ift$},
	ylabel={MSE under attack},
	x label style={at={(axis description cs:0.5,-0.1)},anchor=north},
	y label style={at={(axis description cs:-0.1,.5)},anchor=south},
	xticklabel style = {font=\footnotesize},
	yticklabel style = {font=\footnotesize},
	axis background/.style={fill=white},
	xmajorgrids,
	ymajorgrids,
	legend style={at={(0.02,0.98)}, anchor=north west, legend cell align=left, align=left, draw=white!15!black, font=\scriptsize}]

		\addplot [color={blue}, line width=1pt, mark=*, mark size=1pt]
		table[row sep={\\}]
		{   
			0 	0.640747\\
			0.1  0.67257 \\
			0.2 0.84568  \\
			0.5  1.15614   \\
			1    1.45888 \\
			1.5 1.5452 \\
			2 1.61301  \\
		}
		;\addlegendentry{Our proposed estimator}
		
		\addplot [color={red}, line width=1pt, mark=*, mark size=1pt]
		table[row sep={\\}]
		{   
			0 0.0839695\\
			0.1  0.122709  \\
			0.2 0.337686  \\
			0.5 1.48945\\
			1  4.92193 \\
			1.5 10.9045 \\
			2   17.3378\\
		}
		;\addlegendentry{Kalman estimator}

	
\end{axis}

\end{tikzpicture}